\documentclass[fleqn,usenatbib]{mnras}

\usepackage{newtxtext,newtxmath}

\usepackage[T1]{fontenc}
\usepackage{ae,aecompl}

\usepackage{graphicx}	
\usepackage{amsmath}	
\usepackage{amssymb}	

\title[Periodic orbits of the retrograde coorbital problem]{Periodic orbits of the retrograde coorbital problem}

\author[]{
M.H.M. Morais$^{1}$\thanks{E-mail: helena.morais@rc.unesp.br (MHMM)}
F. Namouni,$^{2}$
\\
$^{1}$Universidade Estadual Paulista (UNESP), Instituto de Geoci\^encias e Ci\^encias Exatas, Av. 24-A, 1515, 13506-900 Rio Claro, SP, Brazil \\
$^{2}$Universit\'e C\^ote d'Azur, CNRS, Observatoire de la C\^ote d'Azur, CS 24229, 06304 Nice, France
}

\date{Accepted XXX. Received YYY; in original form ZZZ}

\pubyear{2019}

\begin{document}
\label{firstpage}
\pagerange{\pageref{firstpage}--\pageref{lastpage}}
\maketitle

\begin{abstract}
Asteroid (514107) Ka`epaoka`awela is the first example of an object in the 1/1 mean motion resonance with Jupiter with retrograde motion around the Sun.  Its  orbit was shown to be stable over the age of the Solar System which implies that it must have been captured from another star when the Sun was still in its birth cluster.  Ka`epaoka`awela orbit is also located at the peak of the capture probability in the coorbital resonance.  Identifying the periodic orbits that  Ka`epaoka`awela and similar asteroids followed during their evolution is an important  step towards precisely understanding their capture mechanism. Here, we find the families of periodic orbits in the two-dimensional retrograde coorbital problem and analyze their stability and bifurcations into three-dimensional periodic orbits. Our results  explain the radical differences observed in  2D and 3D coorbital capture simulations.  In particular, we find that analytical and numerical results obtained for planar motion are not always valid at  infinitesimal deviations from the plane.
\end{abstract}




\section{Introduction}

The Solar System contains only one known asteroid in coorbital resonance with a planet, Jupiter, that moves with a retrograde motion around the Sun: asteroid (514107) Ka`epaoka`awela \citep{Wiegert17,MoraisNamouni17}. Large scale numerical integrations of its past orbital evolution, including perturbations from the four giant planets and the Galactic tide, have shown that it has been at its current location since the end of planet formation 4.5 Gyr in the past. Since a retrograde orbit could not have formed from the material of the Sun's protoplanetary disk at that early epoch, Ka`epaoka`awela  must  have been belonged to a different  star system and was captured by our own when the Sun was still in its birth cluster \citep{NamouniMorais2018}.  Ka`epaoka`awela  is thus the first known example of an interstellar long-term resident in the Solar system.  Understanding exactly how it reached its current location is particularly important.

Coorbital retrograde resonance has been studied in the framework of the restricted three-body problem. Several stable planar and three-dimensional coorbital configurations or modes are known to exist \citep{MoraisNamouni13a,MoraisNamouni16}. Simulations of retrograde asteroids radially and adiabatically drifting towards Jupiter's orbit showed that when retrograde motion is almost coplanar, capture occurs in the coorbital mode that corresponds to Ka`epaoka`awela's current orbit \citep{MoraisNamouni16,NamouniMorais17c}. If motion is exactly coplanar, capture occurs in a distinct coorbital mode \citep{MoraisNamouni16,NamouniMorais17c}. In order to understand such differences and characterize the path that Ka`epaoka`awela  followed in its capture by Jupiter, we aim to identify the periodic orbits of retrograde coorbital motion in the three-body problem.

The importance  of periodic orbits (POs) in the study of a dynamical system has been recognized since the seminal work of Poincar\'{e}. Stable  POs are surrounded by islands of regular (quasiperiodic) motion, whereas chaos appears at the location of unstable POs  \citep{Hadji2006book}.
In the three-body problem, POs are the solutions such that the relative distances between the bodies repeat over a period $T$ \citep{Henon1974}.  They form continuous families and may be classified as (linearly) stable or unstable \citep{Henon1974,Hadji2006book}.  
In the circular restricted three-body problem (CR3BP) with a dominant central  mass, these families may be resonant or non-resonant. The former correspond to commensurabilities between the orbital frequencies, whereas the  latter correspond to circular solutions of the unperturbed (two-body) problem \citep{Hadji2006book}.

In this article, we report on our search of periodic orbits in the CR3BP with a mass ratio $\mu=10^{-3}$.  In Section 2, we explain how we compute the families of periodic orbits and study their stability. In Section 3,  we  describe the families that exist  in the 2D-configuration and  the bifurcations from planar families to the 3D-configuration. In Section 4,  we discuss  how these results  explain the  differences observed in the 2D and 3D capture simulations.  The conclusions of this study  are presented in Section 5.

\section{Computation of periodic orbits}

From the periodicity theorem of \cite{RoyOvenden1955}, symmetric periodic orbits (SPOs) in the n-body problem must fullfill two mirror configurations.  In the CR3BP the possible mirror configurations are: (a)  perpendicular intersection of  the $(x,z)$ plane; (b) perpendicular intersection  of the $x$ axis.  The SPOs may be classified according to the combinations of mirror configurations:  (a)-(a);  (b)-(b); (a)-(b)  \citep{ZagourasMarkellos1977}.   Planar SPOs  intersect the $x$-axis perpendicularly at  times $T/2$ and $T$ \citep{Hadji2006book}.

\cite{MoraisNamouni13a} showed that planar periodic orbits associated with the retrograde 1/1 resonance are symmetric and have multiplicity 2, i.e. they intersect the surface of section $y=0$, perpendicularly ($\dot{x}=0$) and with the same sign for $\dot{y}$, at $t=T/2$ and $t=T$. We use the following standard algorithm to find planar POs: 
\begin{enumerate}
\item A guess initial condition $(x_0,0,0,\dot{y}_0)$  is followed until the 2nd intersection with the surface of section occurs within $|y|<\epsilon_0$ at time $T$. 
\item If $|\dot{x}|<\epsilon$ then the initial conditions correspond to a PO with period $T$. A new search is started varying $x_0$ or $\dot{y}_0$. Otherwise, a differential correction is applied to $x_0$ or $\dot{y}_0$ and the procedure is repeated.
\end{enumerate}

The variational equations have the general solution $\bar{\xi(t)}=\Delta(t) \bar{\xi}(0)$ where $\bar{\xi}(t)$ is the phase-space displacement vector at time $t$ and   $\Delta (t)$ is the 6-dimensional state transition matrix.
The eigenvalues of  $\Delta(T)$ indicate if the PO with period $T$ is linearly stable or unstable.  Due to the symplectic property of the equations of motion  these eigenvalues  appear as reciprocal pairs $\lambda_i \lambda_i^*=1$ ($i=1,2,3$)  and they may be real or complex conjugate. The periodicity condition implies  that one pair of eigenvalues  is $\lambda_1= \lambda_1^*=1$. The PO is linearly  stable if the remaining pairs of eigenvalues are complex conjugate on the unit circle and  unstable otherwise  with instability increasing with the largest eigenvalue's absolute value \citep{Hadji2006book}.  

For planar motion, the state transition matrix, $\Delta_{2}(t)$, is 4 dimensional. The horizontal stability index is $k_{2}=\lambda_2+\lambda_2^*$ where $\lambda_2$ and $\lambda_2^*$ are the  non trivial eigenvalues of $\Delta_{2}(T)$. Stable planar periodic orbits have $-2<k_{2}<2$.
Change of stability occurs when $|k_2|=2$ (or $\lambda_2=\lambda_2^*=\pm1$)  which is often associated with bifurcation of a new family of POs.

The variational equations for displacements out of the plane and in the plane of motion  are decoupled when $z=\dot{z}=0$ \citep{Henon1973} hence  the evolution of $\bar{\xi}_z(t)=(dz,d\dot{z})$ is described by a 2 dimensional  state transition matrix, $\Delta_3(t)$.
The vertical stability index is  $k_{3}=\lambda_3+\lambda_3^*$
where $\lambda_3$ and $\lambda_3^*$ are the eigenvalues  of $\Delta_3(T)$.  Motion  around stable 2D periodic orbits is maintained when there are small deviations out of the plane only if the vertical stability index $-2<k_{3}<2$. When this stability index reaches the critical value 2 (vertical critical orbit or {\it vco}) a bifurcation into a new family of  3D periodic orbits with the same multiplicity may occur \citep{Henon1973,Ichtiaroglou1980}.

To find 3D SPOs that bifurcate from {\it vcos} we follow a procedure similar to   \cite{ZagourasMarkellos1977}:
\begin{enumerate}
\item Initial conditions corresponding to mirror configurations: (a): $(x_0,0,z_0,0,\dot{y}_0,0)$; or (b):   $(x_0,0,0_0,0,\dot{y}_0,\dot{z}_0)$; are followed  until the the 2nd intersection with the surface of section occurs within $|y|<\epsilon_0$  at time $T$.   
\item If $|\dot{x}|<\epsilon$ and $|\dot{z}|<\epsilon$ (a) or  $|\dot{x}|<\epsilon$ and $|z|<\epsilon$ (b) then the initial conditions correspond to a PO  with period $T$ and a new search is started. Otherwise, a differential correction is applied to 2 components of the initial condition vector and the procedure is repeated.
\end{enumerate}

For 3D motion, the pairs of eigenvalues  $\lambda_i$,$\lambda_i^*$ ($i=2,3$) of $\Delta(T)$ are the roots of the characteristic polynomial $\lambda^4+\alpha\lambda^3+\beta\lambda^2+\alpha\lambda+1$ with $\alpha=2-Tr(\Delta(T))$, $2\beta=\alpha^2+2-Tr(\Delta(T)^2)$ \citep{BrayGoudas1967}. They are complex conjugate on the unit circle (linear stability) if  $\delta=(\alpha^2-4(\beta-2))>0$ and$|p|=|(\alpha+\sqrt\delta)|/2<2$, $|q|=|(\alpha-\sqrt\delta)|/2<2$ \citep{ZagourasMarkellos1977}.  Change of stability  with possible bifurcation into a new family of POs occurs when  pairs of eigenvalues coallesce on the real axis while complex instability occurs when  they collaesce on the unit circle and then move away from it \citep{Heggie1985}.

The numerical integration of the CR3BP equations of motion and associated variational equations  were done using the Bulirsch-Stoer algorithm  with  per step accuracy $10^{-13}$. Distance and time were scaled by Jupiter's semimajor axis and orbital period. The computations for an individual test particle were stopped when the distance to a massive body was within its physical radius (taken equal to the Sun's and Jupiter's radius). They were also stopped when the heliocentric distance exceeded 3 times Jupiter's semi-major axis.  

The thresholds for deciding if an orbit is periodic were chosen so that the differential correction procedure converges for each specific type of PO.
We used $\epsilon_0=10^{-11}$ and $\epsilon=10^{-10}$ to find planar and 3D POs. In general, lower (sometimes unfeasible) values are necessary to follow unstable families, as expected due to the exponential divergence of solutions close to unstable POs.  To monitor  the POs computations we checked that  $|\Delta(T)|=1$  with at least 11 significant digits. Stability and bifurcation points were  further checked by  explicitly computing the eigenvalues of $\Delta(T)$. Unstable critical motion (near the transition to stability) was confirmed by computing the  chaos indicator MEGNO \citep{MEGNO}.

\section{The 2D families and bifurcations into 3D }

\cite{MoraisNamouni13a}  showed that  the relevant resonant argument for planar retrograde coorbitals in  the CR3BP is $\phi^*=\lambda-\lambda_p-2\omega$ where $\lambda$ and $\omega$ are the test particle's mean longitude and argument of pericenter, $\lambda_p$ is the mean longitude of the planet. There are 3 types of retrograde coorbitals:  mode1 which  corresponds to libration of $\phi^*$ around 0 and occurs at a wide range of eccentricities; modes 2 and 3 which  correspond to libration of $\phi^*$ around $180^\circ$ and occur, respectively, at small eccentricity (mode 3) and large eccentricity (mode 2).  These modes are retrieved in a 2D model for retrograde coorbital resonance based on the averaged Hamiltonian \citep{Huangetal2018AJ}.

\subsection{Planar SPOs}

We show how the families of SPOs associated with mode 1 (Fig.~1) and modes 2 and 3 (Fig.~2) evolve with the Jacobi constant, $C$. 

Mode 1 resonant POs are horizontally stable when  $C>-1.2256$ and  vertically stable when $C>-1.0507$ ($a<1.0380$, $e>0.1125$). The family ends  by collision with the star when $e\approx 1$.

 \begin{figure*}
 \includegraphics[width=\textwidth]{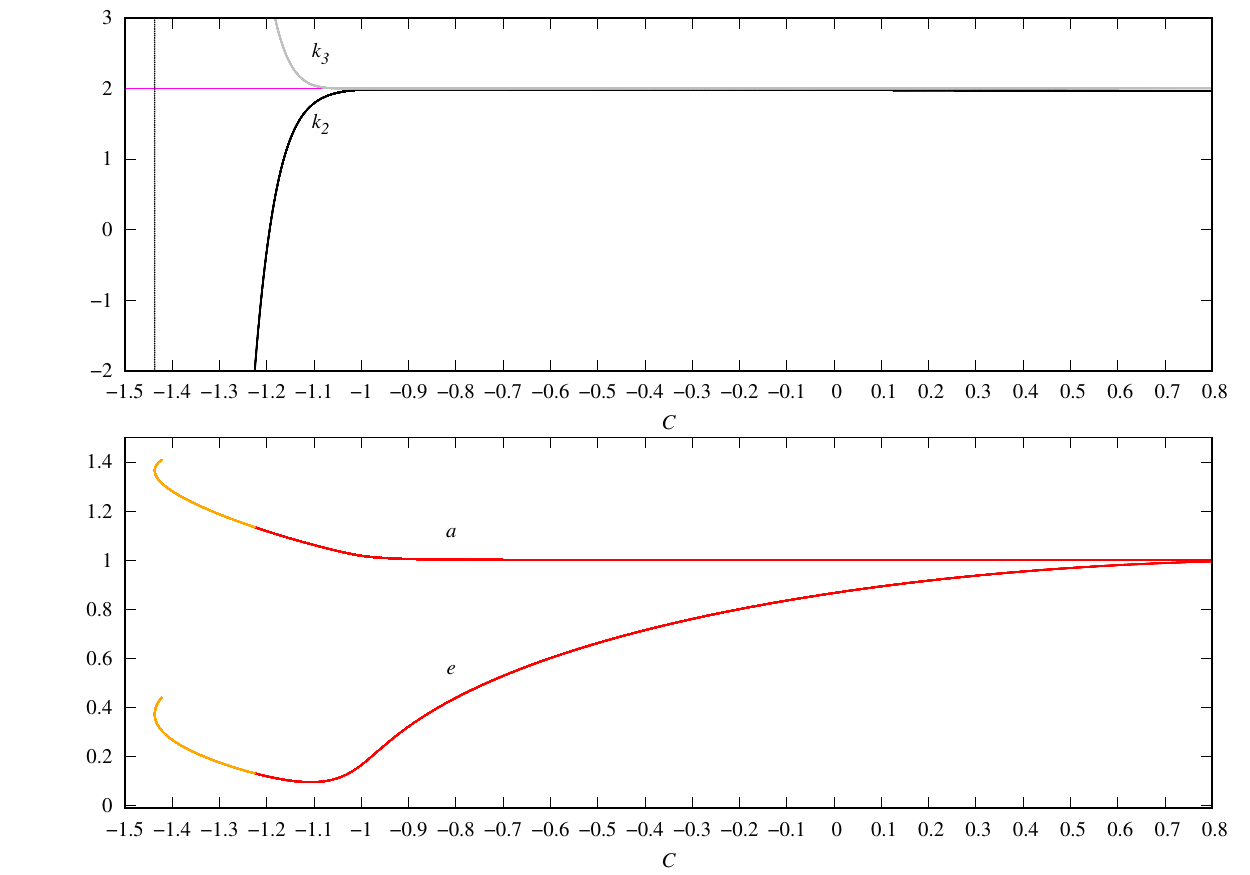}
 \caption{Family of SPOs corresponding to retrograde mode 1 with respect to  the Jacobi constant, $C$.  Top panel: 2D (black) and 3D (gray) stability indexes. Low panel: semi-major axis $a$  and eccentricity $e$  (horizontally stable (red) and unstable (orange)).}
 \end{figure*}

Inner nearly circular non-resonant POs are stable if $C>-0.8429$ ($a<0.9265$). At $C=-0.9562$ ($a=0.9801$) there is a bifurcation into a stable inner resonant PO. This family is stable up to   $C=-0.8643$ ($a=0.9864$, $e=0.3208$: inner mode 3) and stable again from $C=-0.3349$  ($a=0.9976$, $e=0.7411$: mode 2).
Therefore,  mode 2  and inner mode 3 resonant POs  form a single family  which is always vertically stable. The family ends by collision with the star when $e\approx 1$.

Outer  nearly circular non-resonant POs are vertically unstable when $-1.0395>C>-1.1499$ ($1.0218<a<1.0804$). At $C=-1.0387$ ($a=1.0215$) there is a bifurcation into a pair of stable (outer mode 3) and unstable (nearly circular) POs.  Outer mode 3 resonant POs are stable up to $C=-0.9553$ ($a=1.0151$, $e=0.2636$).

  \begin{figure*}
    \includegraphics[width=\textwidth]{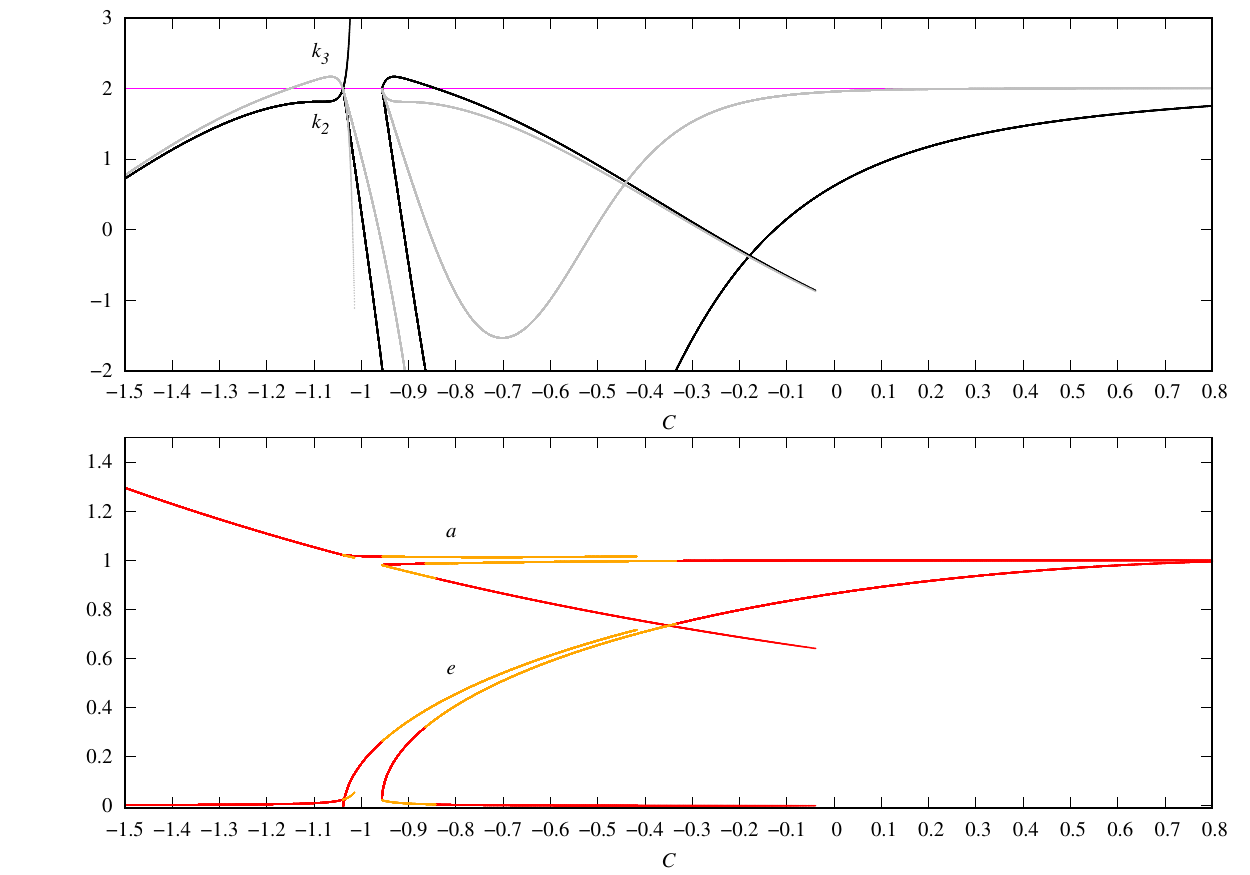}
 \caption{Family of SPOs corresponding to the  outer and inner circular families and  retrograde modes 2 and 3 with respect to  the Jacobi constant, $C$.  Top panel: 2D (black) and 3D (gray) stability indexes. Low panel: semi-major axis $a$  and eccentricity $e$  (horizontally stable (red) and unstable (orange)).}
 \end{figure*}

\subsection{Bifurcations into 3D}

\cite{MoraisNamouni13a,MoraisNamouni16}  showed that in the 3D coorbital problem the relevant resonant angles are $\phi=\lambda-\lambda_p$ and $\phi^*=\lambda-\lambda_p-2\omega$. The 3D retrograde coorbital modes correspond to: $\phi$ librating around $180^\circ$ (mode 4); $\phi^*$ librating around 0 (mode 1) or $180^\circ$ (modes 2 and 3). 

Planar retrograde modes 1 and 2 are horizontally and vertically stable when $C>-1.0507$ and $C>-0.3349$, respectively, hence the associated POs  are surrounded by quasiperiodic orbits in the 3D problem. In particular, quasiperiodic mode 1 and mode 2 orbits may extend down to inclinations $i=90^\circ$  and $i=120^\circ$, respectively \citep{MoraisNamouni16}.

The vertical  critical orbits ({\it vcos}) occur:  on mode 1 family at  $C=-1.0507$ (Fig.~1);   on the outer circular family at $C=-1.1499$ and $C=-1.0395$ (Fig.~2). At the {\it vcos} there are bifurcations into new families of 3D periodic orbits which  we show  in Fig.~3.
 The  mode 2, mode 3 outer and  inner families have no {\it vcos} as the vertical stability index, $k_3<2$. 

At $C=-1.1499$ there is a bifurcation of a nearly circular 2D outer PO into a  stable 3D  resonant PO on configuration (b) which corresponds to  mode 4 
(libration center $\phi=180^\circ$). This family reaches critical stability at  $C= -1.0321$ when $i\approx 173^\circ$. Fig.~4 shows a PO  at this point on the family. Initially,       
 $\phi=180^\circ$ with $\omega$ circulating fast similarly to mode 4 stable branch (Fig.~4: left).  The peaks in $a$ and $e$  occur twice per period, at the encounters with the planet. After $t=8\times 10^3$, chaotic diffusion is obvious  (MEGNO increases linearly with time) and  from $t=1.8\times10^4$ there are transitions between libration around $\omega=90^\circ,270^\circ$  at  small $e$ when $\phi=180^\circ$ and  circulation  around the Kozai centers  $\omega=0,180^\circ$  with eccentricity oscillations up to $e=0.14$ (Fig.~4: right) when $\phi$ circulates.  The Kozai cicles around   $\omega=0,180^\circ$ raise the eccentricity and shift the libration center to $\phi^*=0$.
 
At $C=-1.0507$ there is a bifurcation of planar mode 1 into a  3D PO on configuration (a).  This family is unstable but nearly critical. It has a v-shape with lower / upper branches corresponding the intersection with the surface of section at the apocentric  / pericentric encounters (Fig.~6: left).  There is a bifurcation at $C=-1.0321$ coinciding with the bifurcation on the mode 4 family.  Fig.~5 shows a PO at this bifurcation point.  Initially, $\phi=180^\circ$ with $\omega$ circulating fast (Fig.~5: left).  After  $t=7\times 10^3$, chaotic diffusion is again obvious (Fig.~5: right) with the same qualitative behaviour observed in Fig.~4.

A shift of $t=0.25$ between  the time series in Figs.~4,5 (left) causes overlap of  orbital elements. Further inspection shows that they correspond to the same PO of symmetry type (a)-(b) at different intersections with the surface of section (Fig.~6: right). Hence, the 3D families bifurcating from the {\it vcos} at $C=-1.1499$ and $C=-1.0507$ join at $C=-1.0321$  generating a single unstable circular family which could  be continued  to $i\approx 8^\circ$ and $a= 0.999$.  Since instability on this family increases sharply with decreasing inclination the differential correction scheme stops converging preventing further continuation. We suspect that termination occurs at the Lagrangian point L3 when $i=0$ which is further supported by the shape of the last computed PO in the rotating frame.

At $C=-1.0395$, near the end of the stable branch of the 2D nearly circular outer family, there is a bifurcation into an unstable 3D PO  on configuration (a).  This 3D family corresponds to an unstable fixed point of the coorbital resonance Hamiltonian ($\phi=0$ and $\phi^\star=180^\circ$). It could be continued to $i\approx 88^\circ$, $e\approx0.80$ and $a=1.001$  at which point the family is approaching critical stability.
However,
the long integration of the initial conditions that approximate the last computed
periodic orbit (PO) shows that the eccentricity increases sharply towards unity  around $ t=5\times 10^3$  thus leading to collision with the star. The proximity of the collision singularity prevents further continuation of the family.

 \begin{figure*}
    \includegraphics[width=\textwidth]{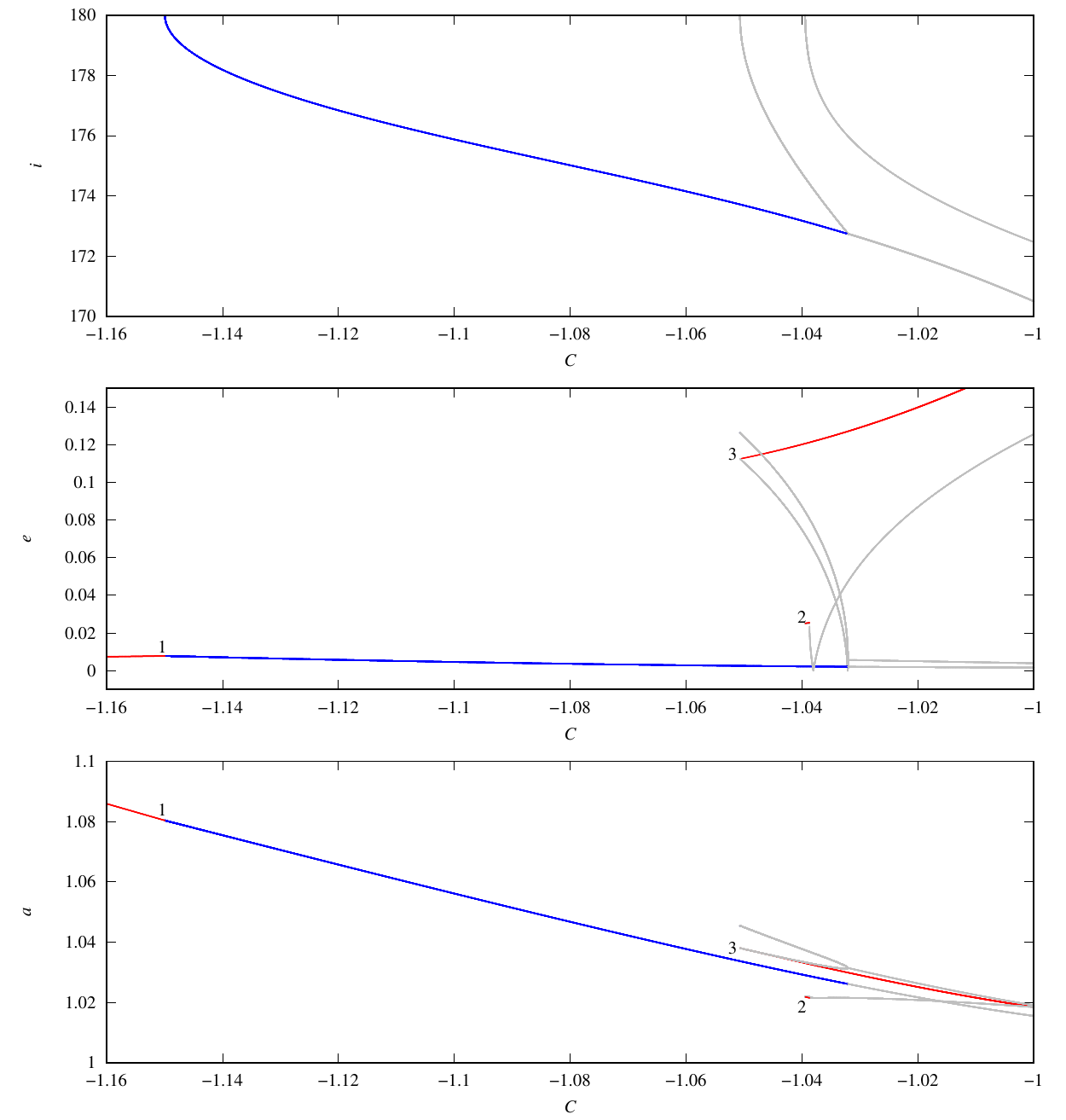}
 \caption{Families of 3D periodic orbits bifurcating from the {\it vcos} at $C=-1.1499$ and $C=-1.0395$ on the outer nearly circular family,
 $C=-1.0507$ on the planar mode 1 family. These  {\it vcos} are labeled 1, 2 and 3, respectively. Top panel: inclination $i$. Mid panel: eccentricity $e$. Low panel: semi-major axis $a$. The families are coloured blue (gray) when stable (unstable). 
 The 2D stable families from which the 3D families bifurcate are coloured red. }
 \end{figure*}
 
 \begin{figure*}
\includegraphics[width=\columnwidth]{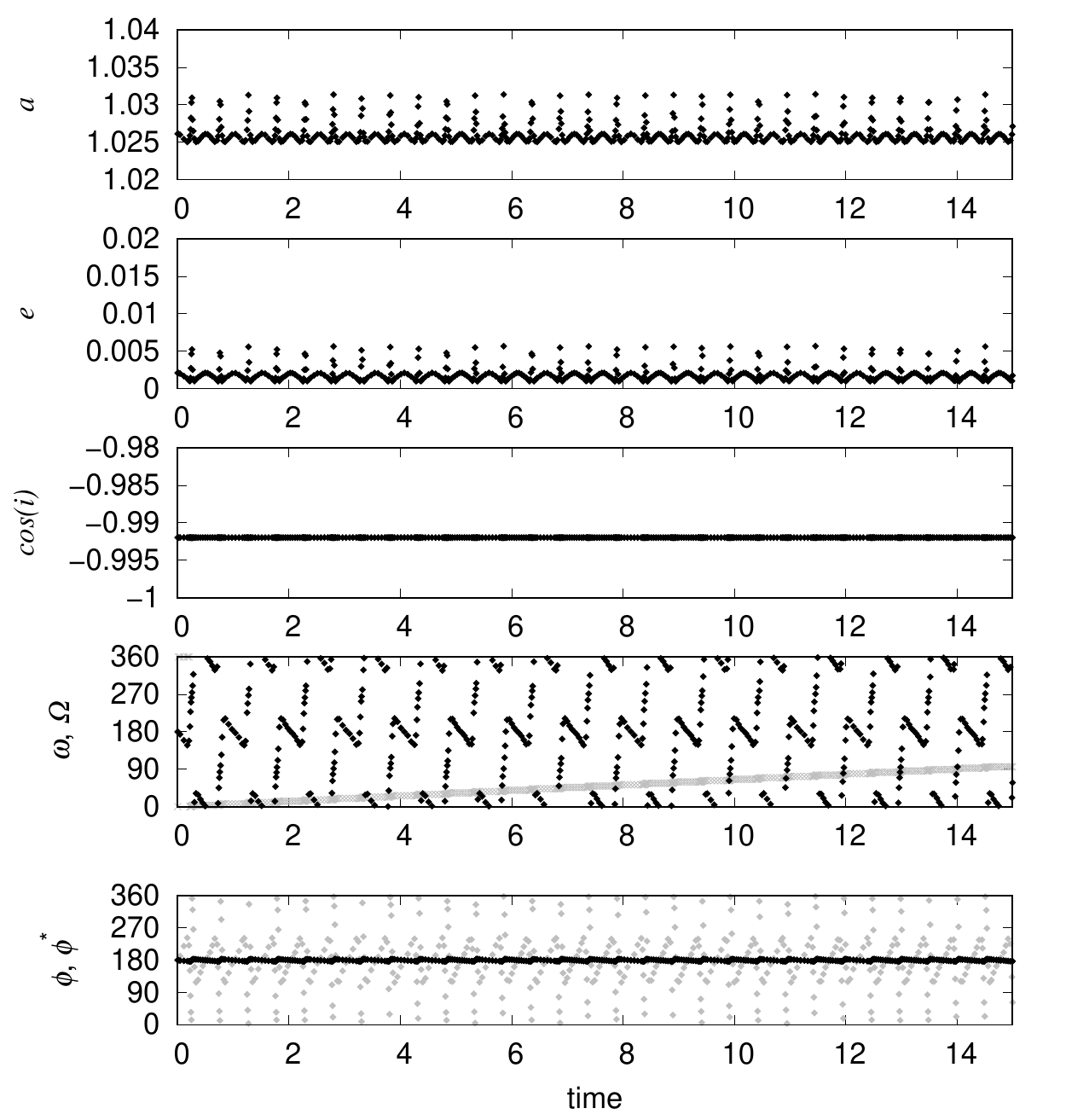}
\includegraphics[width=\columnwidth]{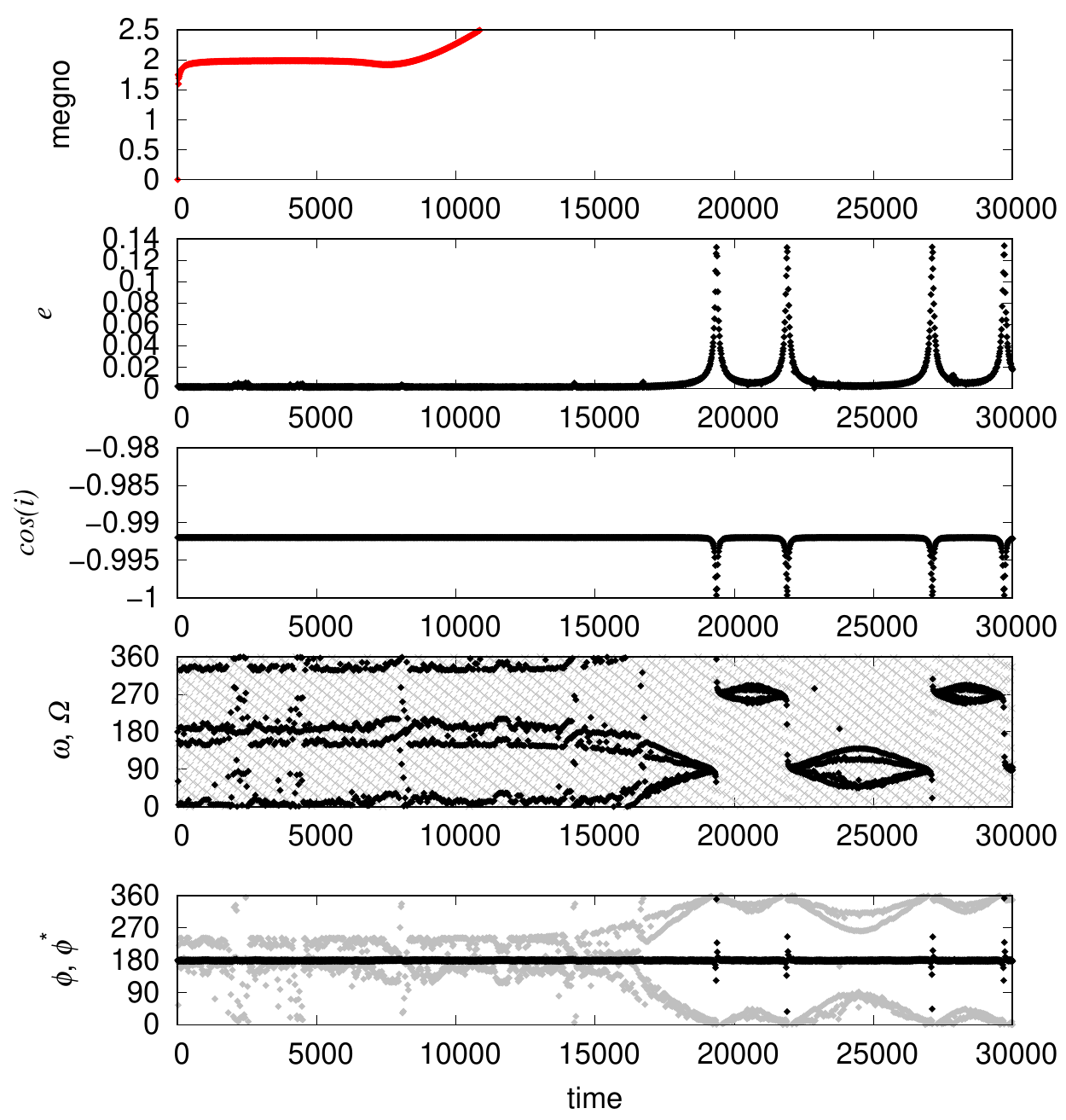}
 \caption{Evolution of PO at the critical point ($C=-1.0321$)  on the 3D mode 4 family. From top to bottom panels:  semi-major axis  / MEGNO; eccentricity, cosine of inclination, argument of pericenter $\omega$ (black) and longitude of ascending node $\Omega$ (gray); resonant angles $\phi$ (black) and $\phi^*$ (gray).}
 \end{figure*}
 
 \begin{figure*}
\includegraphics[width=\columnwidth]{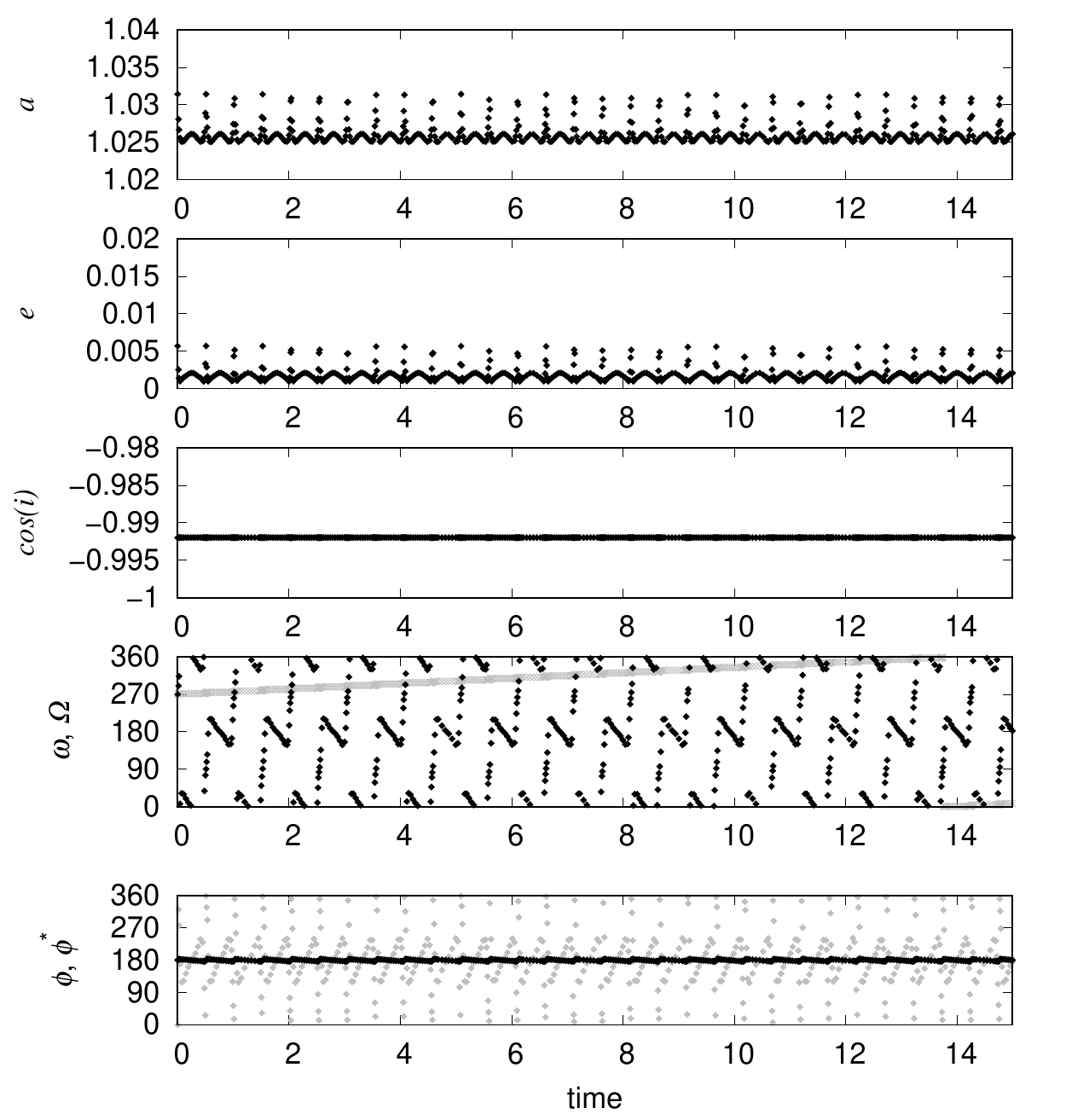}\includegraphics[width=\columnwidth]{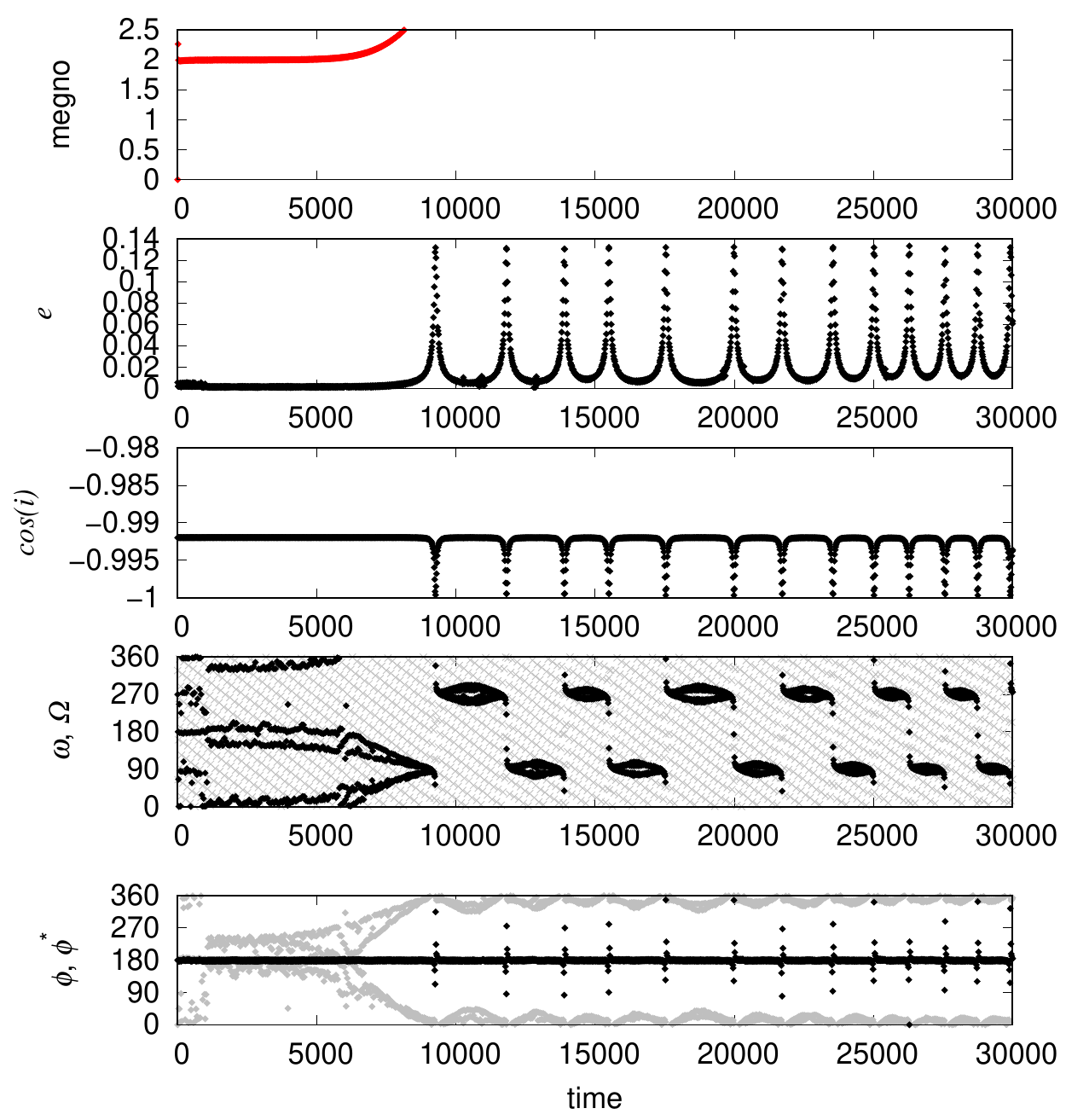}
 \caption{Evolution of PO at the critical point ($C=-1.0321$)  on the 3D family which bifurcates from the {\it vco}  on the planar mode 1 family. Same panels as Fig.~4.}
 \end{figure*}
 
  \begin{figure*}
\includegraphics[width=\columnwidth]{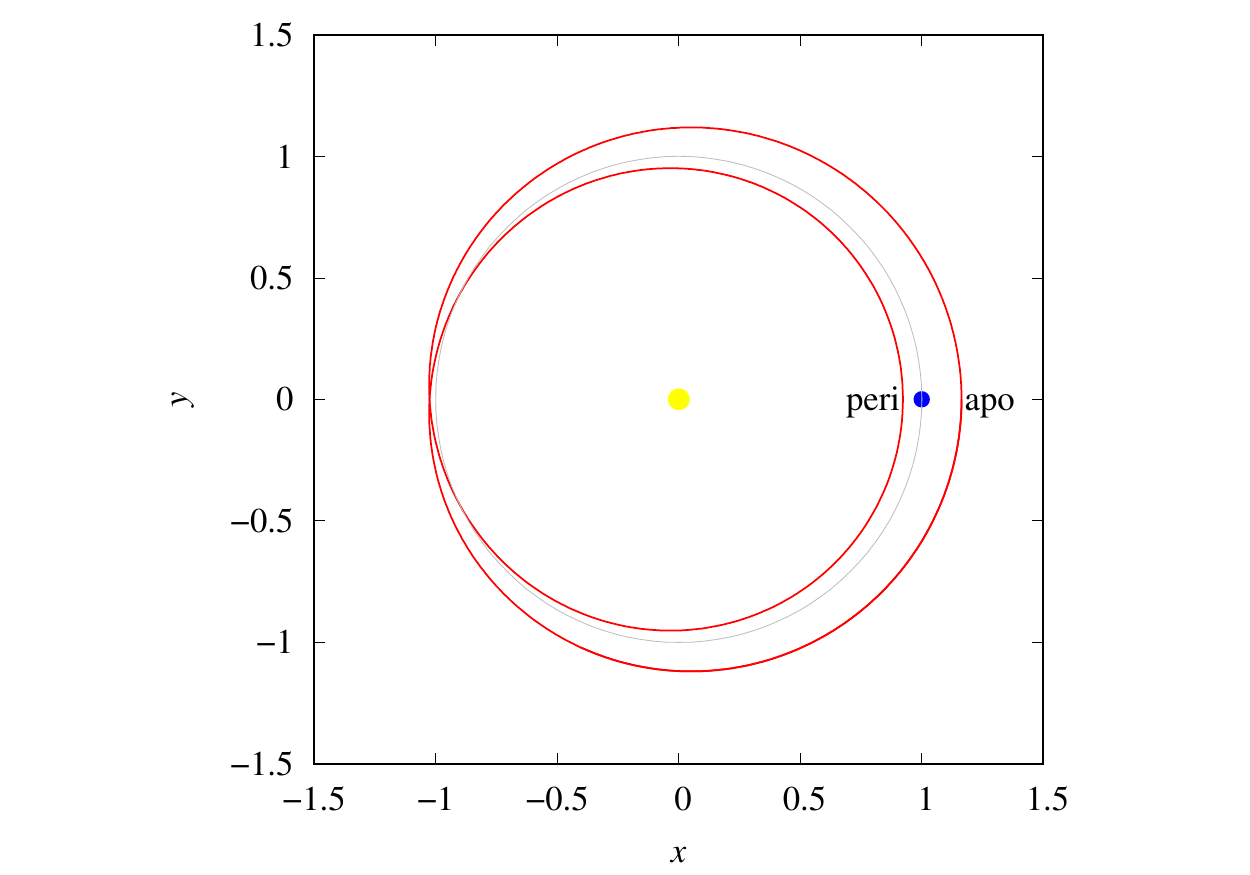}\includegraphics[width=\columnwidth]{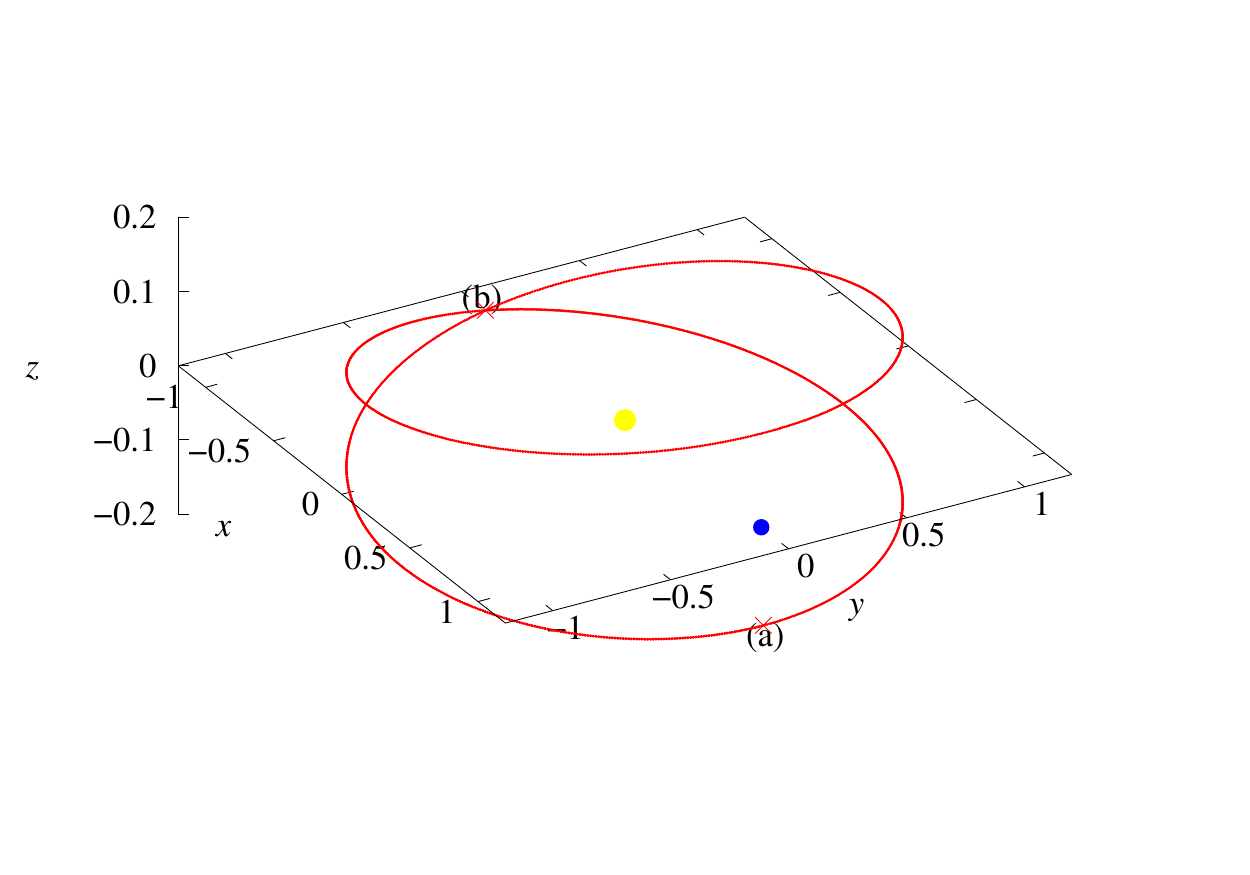}
 \caption{POs in rotating frame:  (left) mode 1 {\it vco}  has encounters and intersections with the surface of section  at pericenter or apocenter; (right)  3D bifurcation at $C=-1.0321$  (mirror configurations  (a) and (b)  are shifted by $t=0.25$).}
 \end{figure*}
 
\section{Coorbital capture in 2D and 3D cases}

In the planar problem, outer  orbits slowly approaching the planet follow the nearly circular non-resonant family which bifurcates into a resonant  SPO at $C=-1.0387$ when $a=1.0215$ (outer mode 3). Capture into outer mode 3 occurs with probability 1 in agreement with \citet{NamouniMorais17c} but the family  becomes unstable at $C=-0.9553$ when $a=1.0151$ and $e=0.2636$.

However, Fig.~3 shows that the behaviour in the (real) 3D problem  at infinitesimal deviations from the plane is radically different.
The nearly circular non-resonant family is vertically unstable between $-1.0395>C>-1.1499$.
The {\it vco} at $C=-1.1499$ ($a=1.0804$) bifurcates into  a resonant mode 4 stable 3D family. 
Hence, outer circular orbits slowly approaching the planet still follow initially the non-resonant family which bifurcates into the 3D mode 4 family.
The inclination then decreases and at $i\approx 173^\circ$ mode 4 family becomes unstable. At this point,  mode 4 family connects with the critical 3D family which bifurcates from the {\it vco} at $C=-1.0507$ ($a=1.0380$, $e=0.1125$) on the stable branch of  the mode 1 planar family. Chaotic transition between the Kozai centers   located at $\omega=90^\circ,270^\circ$ and the separatrices around $\omega=0,180^\circ$ are accompanied by  eccentricity oscillations up to 0.14 and a shift of the libration center towards $\phi^*=0$. 
Exit of this  chaotic region due to a slow decrease in semi-major axis  allows  permanent capture into a quasiperiodic mode 1 orbit, in agreement with the simulations by \cite{MoraisNamouni16,NamouniMorais17c}.
  
Inner circular orbits slowly approaching the planet near the plane become horizontally unstable at $C=-0.8429$ when $a=0.9265$. Therefore, the  inner mode 3 resonant family cannot be reached \citep{MoraisNamouni16}.
The  resonant  family starts at $C=-0.9562$ ($a=0.9801$: inner mode 3),  becomes horizontally unstable at $C=-0.8643$ ($a=0.9864$, $e=0.3208$) and is stable again when $C>-0.3349$ ($a>0.9976$, $e>0.7411$: mode 2).
There are no {\it vcos} on the inner families.

\section{Conclusion}

We showed how the families of periodic orbits for the planar retrograde coorbital problem and their bifurcations into 3D explain the radical differences seen in our capture simulations, namely why 2D orbits are captured into mode 3 while 3D orbits are captured into mode 1 \citep{MoraisNamouni16}. In the planar problem, outer  circular orbits slowly drifting towards the planet follow the  non-resonant family which bifurcates into a resonant mode 3 family at $C=-1.0387$ ($a=1.0215$). This family becomes unstable at $C=-0.9553$ ($a=1.0151$, $e=0.2636$) .  However,  in the (real) 3D problem, mode 3 orbits are never reached.  The nearly circular 2D family becomes vertically unstable ({\it vco}) at $C=-1.1499$ when $a= 1.0804$ where a bifurcation into a 3D resonant family corresponding to mode 4 ($\phi=180^\circ$) occurs. This family becomes unstable when $i\approx 173^\circ$ as it connects  with a 3D family bifurcating from the {\it vco} on mode 1 ( $\phi^*=0$).  Chaotic transitions between the libration centers $\phi=180^\circ$ and $\phi^*=0$ are associated with motion in the vicinity of Kozai separatrices.  As the semi-major axis decreases due to dissipation there is  capture on a mode 1 inclined quasiperiodic orbit, similar to that of Ka`epaoka`awela .  

Our results  explain why mode 1 is the likely end state for objects on retrograde outer circular orbits slowly drifting towards the planet. If the planet migrated inwards, retrograde inner nearly circular orbits become horizontally unstable at $C=-0.8427$ when $a=0.9265$ hence  capture into inner mode 3 is not possible. 
However, eccentric inner retrograde orbits could be captured directly into mode 2  if the relative semi-major axis evolved in discrete steps. Similarly, eccentric outer orbits may be captured directly into mode 1. This could occur e.g.  if the semi-major axis evolves stochastically due to planetary close approaches \citep{Carusi1990}. In the early solar system the latter mechanism (outer eccentric capture) is more likely to occur than the former (inner eccentric capture) and this could explain how  Ka`epaoka`awela  arrived at the current location.

Analytical and numerical results obtained in 2D models are often thought to be valid when the motion is almost  coplanar. Here, we showed  that such extrapolation is not valid for the retrograde coorbital problem. This is due to the vertical instability of the nearly circular 2D family of POs.   A similar mechanism has been reported  for the 2/1 and 3/1 prograde resonances in the planetary (non-restricted)   3-body  problem \citep{Voyatzisetal2014}. 

 Searches for 3D POs typically show that  families end by collision with one of the massive bodies or otherwise exist over the entire inclination range $0 \le i\le 180^\circ$ \citep{Kotoulas&Voyatzis2005,Antoniadou&Libert2019}. Here, we computed the families that originate at the {\it vcos} of the planar retrograde coorbital problem. The family corresponding to an unstable fixed point of the coorbital Hamiltonian ($\phi=0$, $\phi^\star=180^\circ$)  could be continued until it becomes a nearly polar orbit in the vicinity of an instability that leads to collision with the star. The unstable  
doubly-symmetric circular  family corresponding to the libration center $\phi=180^\circ$   seems to end at  the colinear Lagrangian point L3. However, as this family becomes increasingly unstable as the inclination approaches zero its exact termination could not be ascertained.

\section*{Acknowledgements}

Bibliography access was provided by CAPES-Brazil.
M.H.M. Morais research had financial support from  S\~ao Paulo Research Foundation (FAPESP/2018/08620-1) and CNPQ-Brazil (PQ2/304037/2018-4) . 




\bibliographystyle{mnras}
\bibliography{biblio} 

\begin{thebibliography}{}
\makeatletter
\relax
\def\mn@urlcharsother{\let\do\@makeother \do\$\do\&\do\#\do\^\do\_\do\%\do\~}
\def\mn@doi{\begingroup\mn@urlcharsother \@ifnextchar [ {\mn@doi@}
  {\mn@doi@[]}}
\def\mn@doi@[#1]#2{\def\@tempa{#1}\ifx\@tempa\@empty \href
  {http://dx.doi.org/#2} {doi:#2}\else \href {http://dx.doi.org/#2} {#1}\fi
  \endgroup}
\def\mn@eprint#1#2{\mn@eprint@#1:#2::\@nil}
\def\mn@eprint@arXiv#1{\href {http://arxiv.org/abs/#1} {{\tt arXiv:#1}}}
\def\mn@eprint@dblp#1{\href {http://dblp.uni-trier.de/rec/bibtex/#1.xml}
  {dblp:#1}}
\def\mn@eprint@#1:#2:#3:#4\@nil{\def\@tempa {#1}\def\@tempb {#2}\def\@tempc
  {#3}\ifx \@tempc \@empty \let \@tempc \@tempb \let \@tempb \@tempa \fi \ifx
  \@tempb \@empty \def\@tempb {arXiv}\fi \@ifundefined
  {mn@eprint@\@tempb}{\@tempb:\@tempc}{\expandafter \expandafter \csname
  mn@eprint@\@tempb\endcsname \expandafter{\@tempc}}}

\bibitem[\protect\citeauthoryear{{Antoniadou} \& {Libert}}{{Antoniadou} \&
  {Libert}}{2019}]{Antoniadou&Libert2019}
{Antoniadou} K.~I.,  {Libert} A.-S.,  2019, \mn@doi [\mnras]
  {10.1093/mnras/sty3195}, \href
  {https://ui.adsabs.harvard.edu/abs/2019MNRAS.483.2923A} {483, 2923}

\bibitem[\protect\citeauthoryear{{Bray} \& {Goudas}}{{Bray} \&
  {Goudas}}{1967}]{BrayGoudas1967}
{Bray} T.~A.,  {Goudas} C.~L.,  1967, \mn@doi [\aj] {10.1086/110218}, \href
  {https://ui.adsabs.harvard.edu/abs/1967AJ.....72..202B} {72, 202}

\bibitem[\protect\citeauthoryear{{Carusi}, {Valsecchi}  \&
  {Greenberg}}{{Carusi} et~al.}{1990}]{Carusi1990}
{Carusi} A.,  {Valsecchi} G.~B.,   {Greenberg} R.,  1990, \mn@doi [Celestial
  Mechanics and Dynamical Astronomy] {10.1007/BF00050709}, \href
  {https://ui.adsabs.harvard.edu/abs/1990CeMDA..49..111C} {49, 111}

\bibitem[\protect\citeauthoryear{{Cincotta} \& {Giordano}}{{Cincotta} \&
  {Giordano}}{2006}]{MEGNO}
{Cincotta} P.~M.,  {Giordano} M.,  2006

\bibitem[\protect\citeauthoryear{{Hadjedemetriou}}{{Hadjedemetriou}}{2006}]{Hadji2006book}
{Hadjedemetriou} J.~D.,  2006, \href
  {https://ui.adsabs.harvard.edu/abs/2006cwod.book...43H} {p.~43}

\bibitem[\protect\citeauthoryear{{Heggie}}{{Heggie}}{1985}]{Heggie1985}
{Heggie} D.~C.,  1985, \mn@doi [Celestial Mechanics] {10.1007/BF01227832},
  \href {https://ui.adsabs.harvard.edu/abs/1985CeMec..35..357H} {35, 357}

\bibitem[\protect\citeauthoryear{{H{\'e}non}}{{H{\'e}non}}{1973}]{Henon1973}
{H{\'e}non} M.,  1973, \mn@doi [Celestial Mechanics] {10.1007/BF01231427},
  \href {https://ui.adsabs.harvard.edu/abs/1973CeMec...8..269H} {8, 269}

\bibitem[\protect\citeauthoryear{{Henon}}{{Henon}}{1974}]{Henon1974}
{Henon} M.,  1974, \mn@doi [Celestial Mechanics] {10.1007/BF01586865}, \href
  {https://ui.adsabs.harvard.edu/abs/1974CeMec..10..375H} {10, 375}

\bibitem[\protect\citeauthoryear{{Huang}, {Li}, {Li}  \& {Gong}}{{Huang}
  et~al.}{2018}]{Huangetal2018AJ}
{Huang} Y.,  {Li} M.,  {Li} J.,   {Gong} S.,  2018, \mn@doi [\aj]
  {10.3847/1538-3881/aac1bc}, \href
  {https://ui.adsabs.harvard.edu/abs/2018AJ....155..262H} {155, 262}

\bibitem[\protect\citeauthoryear{{Ichtiaroglou} \&
  {Michalodimitrakis}}{{Ichtiaroglou} \&
  {Michalodimitrakis}}{1980}]{Ichtiaroglou1980}
{Ichtiaroglou} S.,  {Michalodimitrakis} M.,  1980, \aap, \href
  {https://ui.adsabs.harvard.edu/abs/1980A&A....81...30I} {81, 30}

\bibitem[\protect\citeauthoryear{{Kotoulas} \& {Voyatzis}}{{Kotoulas} \&
  {Voyatzis}}{2005}]{Kotoulas&Voyatzis2005}
{Kotoulas} T.~A.,  {Voyatzis} G.,  2005, \mn@doi [\aap]
  {10.1051/0004-6361:20052980}, \href
  {https://ui.adsabs.harvard.edu/abs/2005A&A...441..807K} {441, 807}

\bibitem[\protect\citeauthoryear{{Morais} \& {Namouni}}{{Morais} \&
  {Namouni}}{2013}]{MoraisNamouni13a}
{Morais} M.~H.~M.,  {Namouni} F.,  2013, \mn@doi [Celestial Mechanics and
  Dynamical Astronomy] {10.1007/s10569-013-9519-2}, \href
  {http://adsabs.harvard.edu/abs/2013CeMDA.117..405M} {117, 405}

\bibitem[\protect\citeauthoryear{{Morais} \& {Namouni}}{{Morais} \&
  {Namouni}}{2016}]{MoraisNamouni16}
{Morais} M.~H.~M.,  {Namouni} F.,  2016, \mn@doi [Celestial Mechanics and
  Dynamical Astronomy] {10.1007/s10569-016-9674-3}, \href
  {http://adsabs.harvard.edu/abs/2016CeMDA.125...91M} {125, 91}

\bibitem[\protect\citeauthoryear{{Morais} \& {Namouni}}{{Morais} \&
  {Namouni}}{2017}]{MoraisNamouni17}
{Morais} M.~H.~M.,  {Namouni} F.,  2017, Nature, 543, 635

\bibitem[\protect\citeauthoryear{{Namouni} \& {Morais}}{{Namouni} \&
  {Morais}}{2018a}]{NamouniMorais17c}
{Namouni} F.,  {Morais} M.~H.~M.,  2018a, \mn@doi [J. Comp. App. Math.]
  {10.1007/s40314-017-0489-y}, 37, 65

\bibitem[\protect\citeauthoryear{{Namouni} \& {Morais}}{{Namouni} \&
  {Morais}}{2018b}]{NamouniMorais2018}
{Namouni} F.,  {Morais} M.~H.~M.,  2018b, \mn@doi [\mnras]
  {10.1093/mnrasl/sly057}, \href
  {https://ui.adsabs.harvard.edu/abs/2018MNRAS.477L.117N} {477, L117}

\bibitem[\protect\citeauthoryear{{Roy} \& {Ovenden}}{{Roy} \&
  {Ovenden}}{1955}]{RoyOvenden1955}
{Roy} A.~E.,  {Ovenden} M.~W.,  1955, \mn@doi [\mnras]
  {10.1093/mnras/115.3.296}, \href
  {https://ui.adsabs.harvard.edu/abs/1955MNRAS.115..296R} {115, 296}

\bibitem[\protect\citeauthoryear{{Voyatzis}, {Antoniadou}  \&
  {Tsiganis}}{{Voyatzis} et~al.}{2014}]{Voyatzisetal2014}
{Voyatzis} G.,  {Antoniadou} K.~I.,   {Tsiganis} K.,  2014, \mn@doi [Celestial
  Mechanics and Dynamical Astronomy] {10.1007/s10569-014-9566-3}, \href
  {https://ui.adsabs.harvard.edu/abs/2014CeMDA.119..221V} {119, 221}

\bibitem[\protect\citeauthoryear{{Wiegert}, {Connors}  \& {Veillet}}{{Wiegert}
  et~al.}{2017}]{Wiegert17}
{Wiegert} P.,  {Connors} M.,   {Veillet} C.,  2017, Nature, 543, 687

\bibitem[\protect\citeauthoryear{{Zagouras} \& {Markellos}}{{Zagouras} \&
  {Markellos}}{1977}]{ZagourasMarkellos1977}
{Zagouras} C.,  {Markellos} V.~V.,  1977, \aap, \href
  {https://ui.adsabs.harvard.edu/abs/1977A&A....59...79Z} {59, 79}

\makeatother
\end{thebibliography}

\bsp	
\label{lastpage}
\end{document}